# Study on electromagnetically induced transparency effects in Dirac and VO$_2$ hybrid material structure[*]

**DI KE[1,3], XIE MENG[1,3]，XIA HUA RONG[1,3], CHENG AN YU[2], LIU YU[1,3], DU JIA JIA[1,3]\*\***

1. *Department of Optoelectronic Engineering, Chongqing University of Post and Telecommunications, Chongqing 400065, China*
2. *Department of Automation, Chongqing University of Post and Telecommunications, Chongqing 400065, China*
3. *Chongqing Key Laboratory of Autonomous Navigation and Microsystem, Chongqing University of Posts and Telecommunications, Chongqing 400065, China*

In this paper, we present a metamaterial structure of Dirac and vanadium dioxide and investigate its optical properties using the finite-difference time-domain (FDTD) technique. Using the phase transition feature of vanadium dioxide, the design can realize active tuning of the PIT effect at terahertz frequency, thereby converting from a single PIT to a double PIT. When VO$_2$ is in the insulating state, the structure is symmetric to obtain a single-band PIT effect; When VO$_2$ is in the metallic state, the structure turns asymmetric to realize a dual-band PIT effect. This design provides a reference direction for the design of actively tunable metamaterials. Additionally, it is discovered that the transparent window's resonant frequency and the Dirac material's Fermi level in this structure have a somewhat linear relationship. In addition, the structure achieves superior refractive index sensitivity in the terahertz band, surpassing 1 THz/RIU. Consequently, the concept exhibits encouraging potential for application in refractive index sensors and optical switches.

Electromagnetically induced transparency (EIT) is a quantum interference effect in a three-level atomic system. A "transmission window" with a specific width in the wide absorption spectrum is created by the interference of two distinct quantum transition channels, which cooperate to produce the effect. In 2008, a periodic metamaterial structure consisting of three metal rods was designed[1]. The team observed for the first time an EIT-like effect at room temperature based on this material, which is free from harsh experimental conditions. This kind of EIT-like effect is known as Plasmon induced transparency (PIT). There are two ways to realize the electromagnetic-induced transparency effect[2]: one is the interference between the bright mode that can be directly excited by the incident electromagnetic wave and the dark mode that cannot be directly excited, thus generating a transparent window; The other is through interaction between two bright modes, which weakens the absorption of electromagnetic waves near the resonance frequency to create a transparent window. EIT effect based on metamaterials can be observed in the visible[3], near-infrared[4], terahertz[5], and radio frequency bands[6]. Terahertz waves have a unique spectral position, resulting in higher resolution, lower quantum energy, and stronger penetration than other types of waves[7]. The metamaterials based on terahertz in various applications such as tunneling detection[8], biosensing[9], and medical diagnostics[10]. At the same time, graphene materials have been widely used in tunable PIT metamaterials design in recent years due to the excellent properties of 2D graphene materials, such as electrically tunable[11] and plasmonic loss[12,13]. By adjusting the Fermi energy level, a metamaterial structure made of graphene and metal is suggested to actively modify EIT peak intensity at terahertz frequencies without causing any frequency shift[14]. A metal super surface design with a double EIT effect is proposed in the terahertz regime, and active tuning of the two windows is further realized by controlling the graphene structure[15]. A metal-graphene hybrid metamaterial structure is proposed, and dynamic tuning on the metamaterial slow-light effect is realized by adjusting the graphene Fermi energy level[16]. However, Dirac semimetals have a carrier mobility of up to $9\times10^6 cm^2\cdot V^{-1}\cdot S^{-1}$（5K）[17], which is superior to any semiconductor and about 45 times（$2\times10^5 cm^2\cdot V^{-1}\cdot S^{-1}$，5K）the carrier mobility of graphene compared to graphene[18]. Simultaneously, Dirac materials exhibit exceptionally rapid response times and swift relaxation times, rendering them indispensable in the domains of optics, electronics, and related disciplines.

---

[*] This work has been supported by Young Scientists Fund of the National Natural Science Foundation of China (Grant No. 11704053), the National Natural Science Foundation of China (Grant No. 52175531) and the Science and Technology Research Program of Chongqing Municipal Education Commission (Grant No. KJQN 201800629，KJZD-M202000602, 62375031).

\*\*　E-mail: dujj@cqupt.edu.cn

This research introduces a new metamaterial model inspired by EIT and utilizing Dirac materials. The proposed model exhibits dynamic tunability. The structure's PIT effect in the terahertz band is computed using the FDTD algorithm, showcasing its potential for applications in academic research and beyond. We achieve dynamic tunability of the PIT effect by designing patterned Dirac materials combined with vanadium dioxide materials. The modulation of Dirac conductivity can be achieved by manipulating the Fermi energy level, thereby allowing for the design of optical switches with multiple bands. Furthermore, a comprehensive investigation is conducted to analyze the sensing features of the transparent window in terms of refractive index. This analysis showcases the potential of the proposed design in the field of refractive index sensors.

The model and geometric parameters of the proposed structure are shown in Fig.1. The structure consists of two split ring resonators (SRR) and a cut wire (CW). Two exactly same split rings are placed symmetrically on both sides of the cut wire, the opening length of the split ring is defined as $b$. The specific parameters of the structure are shown in Fig. 1(a). The width and length of the CW are respectively defined as $W_1$=3um and $L_1$=28um. The length of the SRR is $L_2$=19um, the width is $L_3$=10um, and the arm width is $W_2$=2um. The distance between the split ring and the cut wire is defined as the coupling distance $g$, setting $g$=2.5um. The size of the substrate material is 45um and the thickness is $h$=15um. The range of the simulated light source is set to 1 THz-5 THz, and the direction of the electric field is at $\theta$ angle to the X-axis of the metamaterials. The whole metamaterial structure is placed on the substrate with period $P$ and refractive index 1.5 as in Fig.1(b). Both the SRR and CW structures are composed of Dirac material. In the non-zero temperature case, the dynamic conductivity of the Dirac 3D electron gas can be derived from the approximate Kubo equation with the following real and imaginary part values[19]:

$$\text{Re}[\sigma(\Omega)] = \frac{e^2}{\hbar} \frac{gk_F}{24\pi} \Omega G(\Omega/2), \quad (1)$$

$$\text{Im}[\delta(\Omega)] = \frac{e^2}{\hbar} \frac{gk_F}{24\pi^2} \{\frac{4}{\Omega}[1+\frac{\pi^2}{3}(\frac{T'}{E_f})]\} \\ + 8\Omega \int_0^{\varepsilon_c} [\frac{G(\varepsilon)-G(\Omega/2)}{\Omega^2-4\varepsilon^2}]\varepsilon d\varepsilon, \quad (2)$$

Here, $G(E) = n(-E) - n(E)$, and $n(E)$ is Fermi-Dirac distribution. $k_F = E_f / \hbar v_F$ is Fermi momentum, $T'$ is non-zero temperature, $\hbar$ is reduced Planck constant, $v_F = 10^6 \text{ms}^{-1}$ is Fermi speed, $E_f$ is Fermi energy level. Thus, the conductivity of Dirac can be tuned by changing the Fermi energy level. The scattering rate $\Omega$ can be expressed as $\Omega = \hbar\omega/E_f + i\hbar\tau^{-1}/E_f$, and $\varepsilon_c = E_c/E_f$ denoted as the cutoff energy. According to Eq. (1) and Eq. (2), if only considering the longitudinal dynamic conductivity, the local dynamic conductivity of the Dirac material in the low-temperature limit can be approximated as[20]:

$$\text{Re}[\sigma(\Omega)] = \frac{e^2}{\hbar} \frac{gk_F}{24\pi} \Omega\theta(\Omega-2), \quad (3)$$

$$\text{Im}[\delta(\Omega)] = \frac{e^2}{\hbar} \frac{gk_F}{24\pi} [\frac{4}{\Omega} - \Omega In(\frac{4\varepsilon_c^2}{|\Omega^2-4|})], \quad (4)$$

And the dielectric constant can be obtained by the following equation:

$$\varepsilon = \varepsilon_b + i\sigma/\omega\sigma_0, \quad (5)$$

The Dirac model used in this paper is based on AlCuFe quasi-crystals[21], hence, $g = 40$, $\varepsilon_b = 1$.

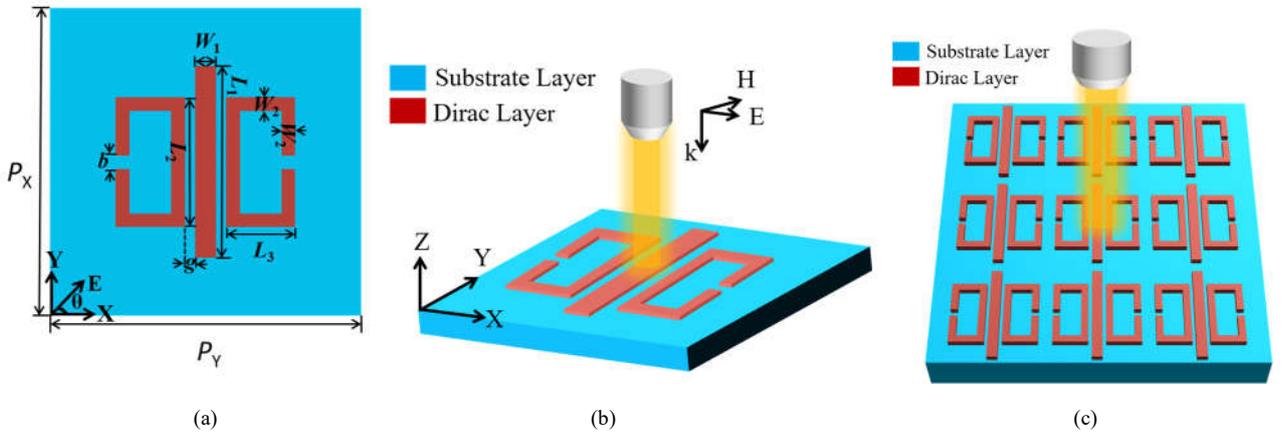

**Fig. 1 Structural model of the single-band PIT. Single-band PIT structure's (a) 2D planar structure and specific parameters (b) 3D structural units (c) 3D spatial modeling**

The design is simulated and validated using the commercial software FDTD Solutions. As shown in Fig.1(c), the model is periodic in both the X and Y directions. In order to achieve an exact simulation, the boundary conditions were defined as periodic boundaries in the X and Y dimensions, while a perfect absorption layer was implemented in the Z direction.

The simulation results of the single band PIT structure are shown in Fig. 2. Fig. 2(a) displays the transmission curves of the SRR structure, CW structure, and SRR-coupled CW when interacting with the incident light. The SRR operates at a frequency of 2.32 THz and has a Lorentzian linear transmission curve, which means that it can interact with incident light and is considered a bright mode. Similarly, the CW structure is a bright mode and has a resonance curve at a frequency of 3.63 THz. When the two structures interact, a bright-bright mode coupling method is formed, resulting in a transparent window at 2.77 THz. This transparent window is labeled as B in Fig. 2(a) with 95% transmittance, and two transmission valleys have been assigned labels A and C, with specific frequencies of 2.31 THz and 3.58 THz, respectively. Fig. 2(b)-(d) show the electric field results at frequencies A, B, and C. Through a comprehensive examination of the distribution pattern, it is possible to ascertain the origin of the transparent window. In Fig. 2(b), it can be observed that the electric field is solely localized on the SRR structure, particularly concentrated near the aperture. Consequently, LC resonance occurs in the SRR, leading to the formation of the transmission valley at A. The CW is excited and resonating in Fig. 2(d), resulting in a significant number of electric fields that converge at the top and bottom of the CW to generate transmission valleys C. The electric field distribution at the transparent window is depicted in Fig. 2(c). It is observed that the electric field strength on the SRR and CW structure is considerably diminished when compared to Fig. 2(b) and Fig. 2(d). This reduction in electric field strength can be attributed to the destructive interference effect between the SRR and CW. Consequently, the transmittance of the structure is enhanced, leading to the formation of a transparent window.

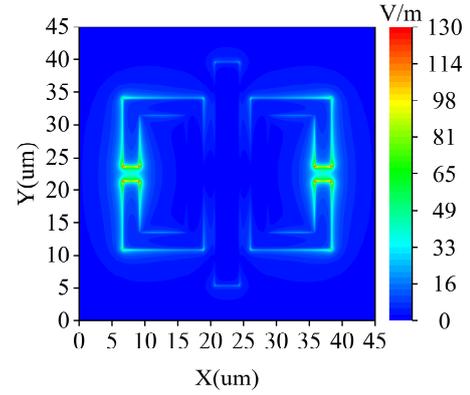

(b)

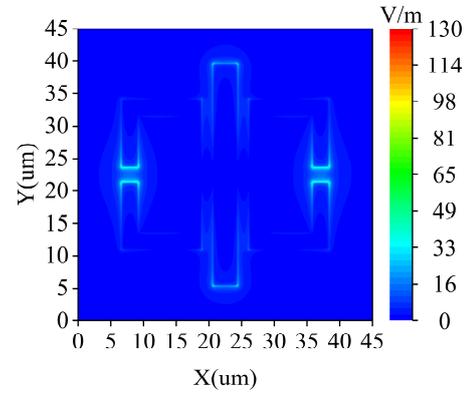

(c)

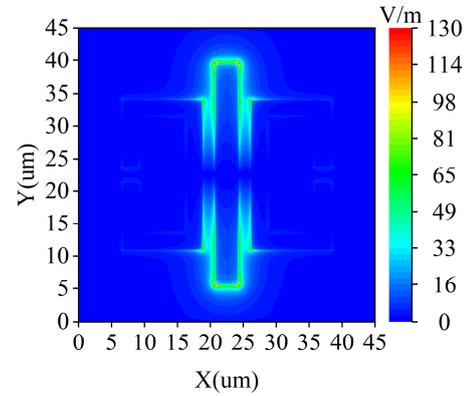

(d)

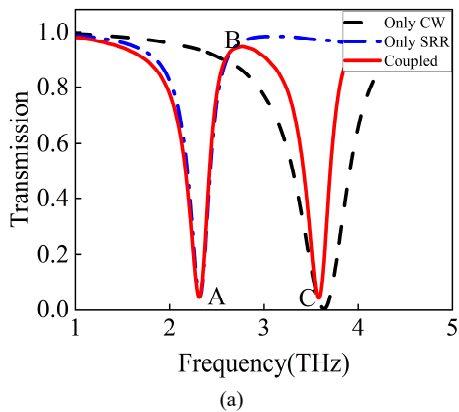

(a)

Fig.2 Transmission spectra and electric field distribution results for single-band structure. (a) Transmission spectrum of SRR and CW resonating individually and SRR coupled CW results. Electric field distribution at the frequency of (b) transmission valley A (c) transmission peak B (d) transmission valley C

In the following section, we present a dual-band photonic integrated circuit (PIT) structure that incorporates vanadium dioxide ($VO_2$) as an auxiliary material, building upon the single-band PIT structure described before. $VO_2$ is a frequently seen phase change material that exhibits metal-insulator phase change characteristics owing to its unique crystal structure[22].

This property enables the material to transition between insulating and metallic state. Therefore, we use VO$_2$ material to fill the opening of the SRR structure in equal proportion and remodel the spilled ring structure into a square ring (SR) structure to introduce a new mode, which breaks the symmetry of the overall structure to obtain a remarkable electromagnetic-induced transparency phenomenon. The specific dual PIT structure model is shown in Fig. 3.

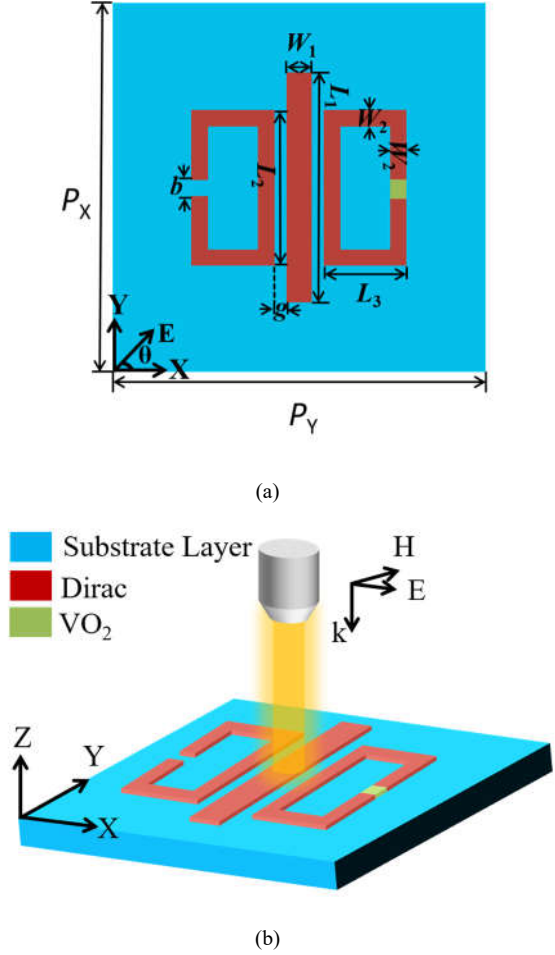

(a)

(b)

**Fig.3 Dual-band PIT structure model. (a) 2D model and specific structural parameters. (b) Three-dimensional model.**

In the terahertz frequency range, the relative permittivity of VO$_2$ can be described by Drude modeling[23]:

$$\varepsilon_{VO_2} = \varepsilon_\infty - \frac{\omega_p^2(\sigma_{VO_2})}{\omega^2 + i\gamma\omega} \quad (6)$$

There, $\varepsilon_\infty = 12$ is the high-frequency dielectric constant, $\gamma = 5.75 \times 10^3$ rad/s is the collision frequency, and $\sigma_{VO_2}$ is the conductivity of vanadium dioxide; the plasma frequency can be described approximately as:

$$\omega_p^2(\sigma_{VO_2}) = \frac{\sigma_{VO_2}(\omega)}{\sigma_0}\omega_p^2(\sigma_0) \quad (7)$$

Here: $\omega_p(\sigma_{VO_2})$ is the value of vanadium dioxide conductivity, and the corresponding plasma frequency when the value is $\sigma_{VO_2}$ taken as $\sigma_0 = 3 \times 10^5$ S/m, $\omega_p(\sigma_0) = 1.4 \times 10^{15}$ rad/s. The conductivity of VO$_2$ can be changed by 3-5 orders of magnitude during the transition of VO$_2$ from the insulating state to the metallic state.

The simulation investigates the variation in conductivity of VO$_2$, ranging from 1000S/m to 200000S/m, to see its transition from an insulating state to a metallic one. The obtained simulation findings are presented in Fig. 4. Fig. 4(a) illustrates that the positions of the transparent window at B and the transmission valleys A and C remain unaltered as the conductivity of VO$_2$ increases. However, a new transparent window is created at 3.86 THz, and the transmission spectrum changes from a single-band PIT to a dual-band PIT. We label the newly generated transparent window and transmission valley as D and E respectively as shown in Fig. 4(a), and the specific frequency of the transmission valley E is 4.41 THz. Fig. 4(b) shows the transmission spectra of each structure in the dual PIT structure at their respective resonances. The SRR and CW structures obtain transmission curves at 2.32 THz and 3.63 THz, respectively which is consistent with the results of the single-band structure, while the square ring (SR) with mixed VO$_2$ material obtains a transmission curve at 4.57 THz. Although the SR structure is also regarded as a bright mode, this mode is introduced by changing the symmetry of the structure and we can control the presence or absence of this bright mode by just controlling the VO$_2$ conductivity. Therefore, the method of controlling transparent windows in this paper is more positive and effective compared to the regulatory method by changing the structural parameters. The transmission spectrum curves in Fig. 4(c) are the result of the two-by-two coupling of the three structures. CW coupled with SRR structure produces a transparent window with higher amplitude at 2.66 THz. CW coupled with SR produces a transparent window with a smaller amplitude at 3.90 THz. The transparent window at location B is a consequence of the coupling between CW and SRR. On the other hand, the window at location D is formed due to the interaction between CW and SR. The transmission spectral window achieved through the coupling of SRR with SR is essentially a combination of the individual transmission spectra of the SRR and SR. This combination does not result in any interference effects.

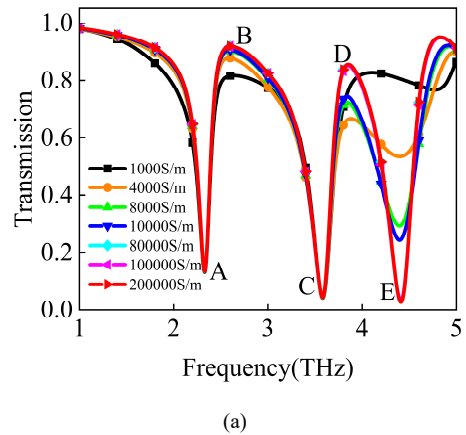

(a)

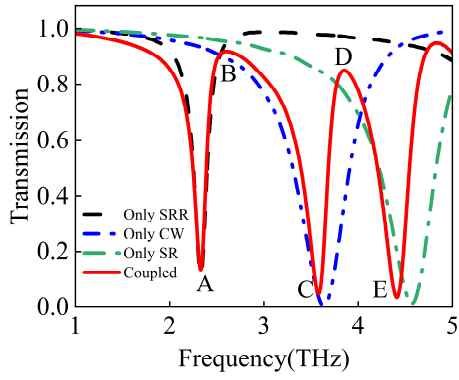

(b)

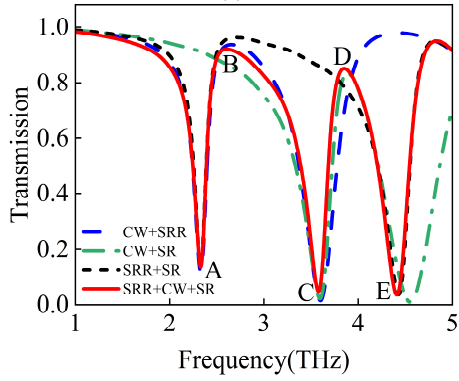

(c)

**Fig.4 Transmission spectrum results of dual-band PIT. (a) VO$_2$ conductivity affects PIT effect result. (b) Transmission spectrum results of SRR, CW and SR structures when they act alone. (c) Transmission spectrum results for two-by-two coupling of CW+SRR, CW+SR and SRR+SR groups.**

Fig. 5 presents the electric fields extracted at the frequencies corresponding to the locations of A, B, C, D, and E, respectively. As shown in Fig. 6(b), the same source as in the single-band PIT structure, the SRR structure as in Fig. 5(a) is in LC resonance at A, the CW as in Fig. 5(c) is in dipole resonance at C, and the transparent window at B arises from destructive interference between the two. Similarly, the generation of the transparent window at D has the same mechanism as the formation of the B window, only the main body of the acting structure is different. At the transmission valley E, the SR structure is in a dipole resonance state, and a large electric field is concentrated on the structure as shown in Fig. 5(e). The transparent window at D originates from the destructive interference between the CW and SR structures, which is reflected in the electric field strengths of the SR and CW at this frequency as shown in Fig. 5(d), which are significantly weaker than the transmission valley C and transmission valley E.

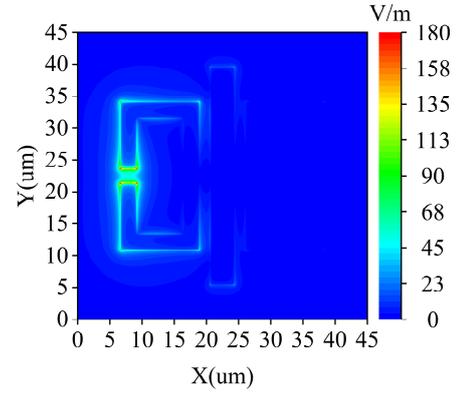

(a)

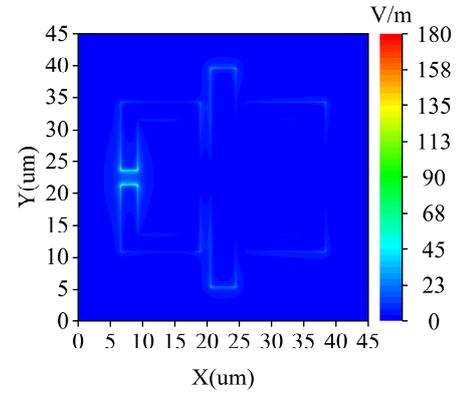

(b)

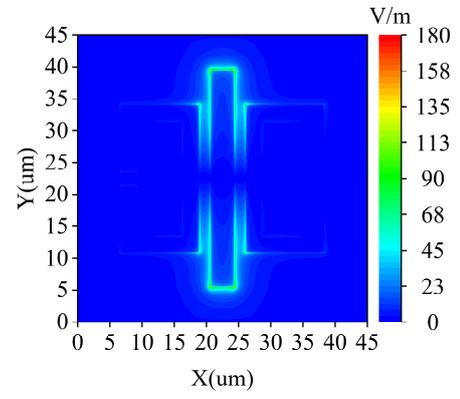

(c)

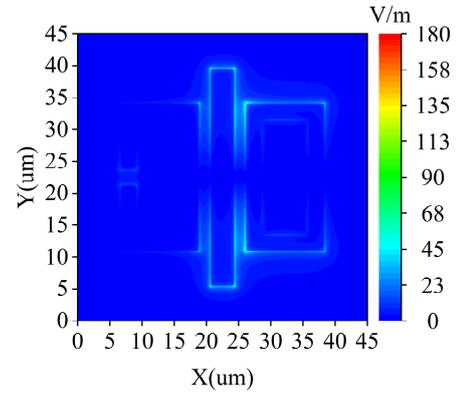

(d)

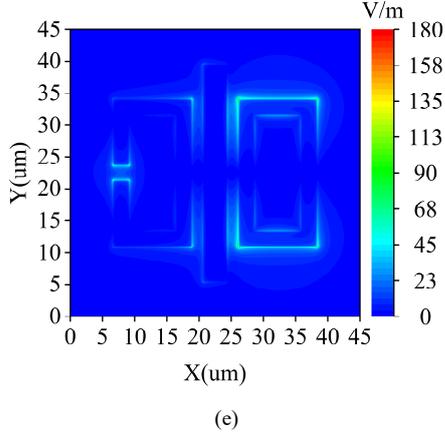

(e)

**Fig. 5 Electric field distribution for the dual PIT effect. Electric field distribution at the frequency of (b) transmission valley A (c) transmission peak B (d) transmission valley C.**

Fig. 6(a) shows how the transmission spectrum changes as the Dirac Fermi energy level changes from 0.10 eV to 0.50 eV. As the Fermi energy level of the Dirac material increases, the transmission spectrum shifts to higher frequencies. When the Fermi energy levels are 0.10 eV and 0.3 eV, respectively, their transparent windows and transmission valley positions overlap. The relationship between the transmission valley and transmission peak at each frequency, as well as their dependence on the Fermi energy level, can be observed in Fig. 6(b). It is evident that the Fermi energy level exhibits a linear correlation with these transmission features. Furthermore, the dual-band PIT effect is found to yield a greater number of frequency points compared to the single-band PIT effect. This exhibits its potential for applications in optical switching. As shown in Fig. 6(c), when the Fermi energy level is 0.10 eV, the structure at 3.25 THz and 4.13 THz are regarded as "off" state; when the Fermi energy level is adjusted to 0.30 eV, the structure becomes "on" state at 3.25 THz and 4.13 THz and "off" state at 2.50 THz and 4.65 THz, thus realizing the switching design at several different frequencies. The amplitude modulation $T_{mod}$ is used to describe the modulation capability of photoelectric switches and is defined as follows[24]:

$$T_{\text{mod}} = \frac{|T_{\text{on}} - T_{\text{off}}|}{T_{\text{on}}} \times 100\% \quad (8)$$

Here: $T_{on}$ and $T_{off}$ are the signal strengths in the "on" and "off" states, respectively. In this paper, they are the transmittance values at the corresponding frequencies. Tab.1 lists parameter results related to the dual PIT structure as an optical switch. The structure achieves switching modulation amplitudes of 96.63%, 84.15%, 91.76% and 96.67% at 2.50 THz, 3.25 THz, 4.13 THz and 4.65 THz, respectively. Meanwhile, the performance of optical switches with metamaterials from different references is compared in Tab.2. It can be seen that the optical switch designed in this paper has a better performance in amplitude modulation, which makes this design valuable for application.

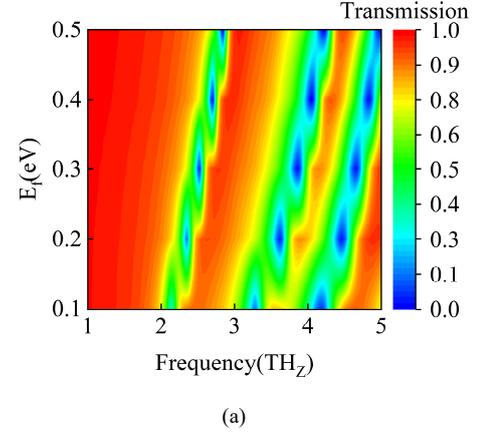

(a)

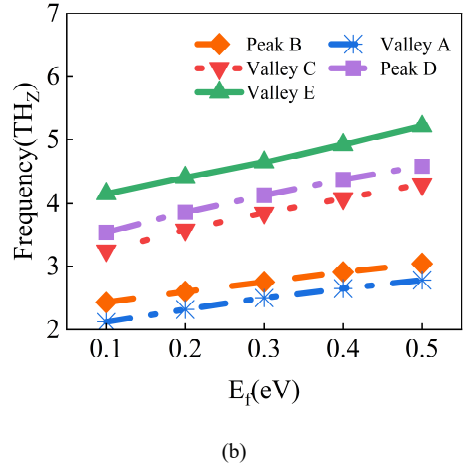

(b)

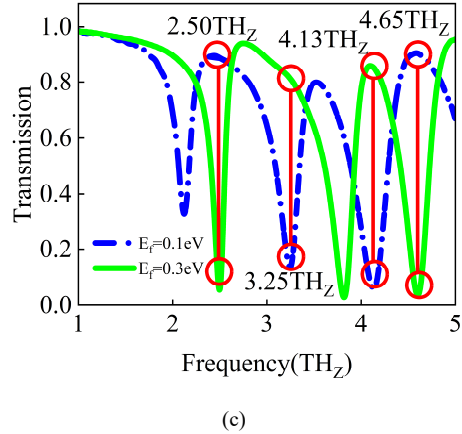

(c)

**Fig. 6 Dual-band PIT varying with different Fermi energy levels. (a)Transmission spectrum varying Transmission spectrum varying with Fermi energy level. (b)Frequency of transmission peaks and transmission valleys versus Fermi energy level. (c)Schematic diagram of the transmission spectrum at 0.1EV.**

**Tab.1 Characteristics of dual-band metamaterial optical switches**

| Frequency(THz) | $E_f$=0.10eV | | $E_f$=0.30eV | | $T_{mod}$ |
|---|---|---|---|---|---|
| | Transmission | on/off | Transmission | on/off | |
| 2.50 | 0.89 | On | 0.03 | Off | 96.63% |
| 3.25 | 0.13 | Off | 0.82 | On | 84.15% |
| 4.13 | 0.07 | Off | 0.85 | On | 91.76% |
| 4.65 | 0.90 | on | 0.03 | off | 96.67% |

**Tab.2 Comparison of optical switching performance in different metamaterials**

| References | Amplitude modulation degree | Year |
|---|---|---|
| [25] | 74.7% 87.8% 76.5% 77.7% | 2020 |
| [26] | 67.5% 86.1% 65.3% | 2021 |
| [27] | 57% 67% 61% | 2022 |
| [28] | 90% 95.6% | 2023 |
| this work | 96.63% 84.15% 91.76% 96.67% | -- |

Metamaterial resonance in the high frequency band can be equal to plasma resonance[29], and its resonant frequency is defined as:

$$\omega_q \propto \frac{1}{2d\sqrt{\varepsilon_{eff}}} = \frac{1}{2d\sqrt{\Omega\varepsilon_{air}+(1-\Omega)\varepsilon_m+\varepsilon_{sub}}}, \quad (9)$$

In the above equation, $d$ is the size of the unit structure, $\varepsilon_{eff}$ denotes the equivalent dielectric constant of the environment surrounding the metamaterial. $\varepsilon_{air}$、$\varepsilon_m$ and $\varepsilon_{sub}$ represent the dielectric constant of air, object to be measured on metamaterial surface, and the substrate material, respectively. The percentage of air in the environment is defined as $\Omega$。From the above equation, the resonant frequency of the metamaterial is inversely proportional to the arithmetic square root of the equivalent dielectric constant of its surroundings. We use sensitivity ($S$) to describe the sensing performance for refractive index[30]：

$$S = \frac{\Delta\lambda}{\Delta n} \quad (10)$$

where $\Delta\lambda$ denotes the amount of frequency shift with environmental refractive index. $S$ indicates the amount of wavelength shift due to a unit change in refractive index. The higher sensitivity indicates a greater refractive index detection capability of the structure. In Fig. 7(a) shows that the transmission spectrum varies with the environmental refractive index without the VO$_2$ material. As the refractive index changes from 1.1 to 1.2, the whole transmission spectrum shifts to lower frequencies, which is consistent with Eq. (9). Fig. 7(b) illustrates the linear relationship between the resonant frequency and the refractive index of transparent window B. The resonant frequency of the transparent window exhibits a strong linear correlation with the refractive index, as evidenced by the good fit of the data points. In this context, the sensitivity to changes in the refractive index of transparent window B is denoted as $S_B$. The slope of the fitted curve, which represents the magnitude of $S_B$, is determined to be 1 THz/RIU.

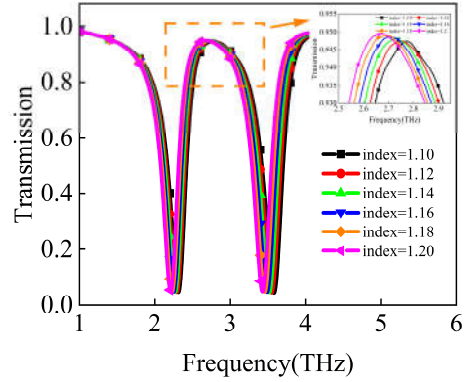

(a)

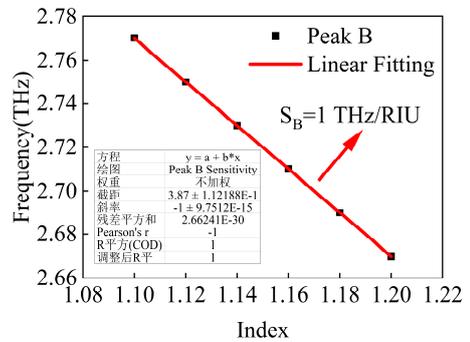

(b)

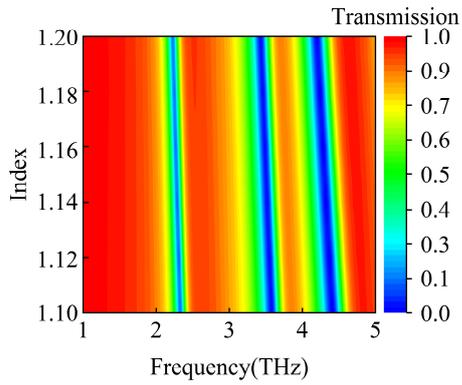

(c)

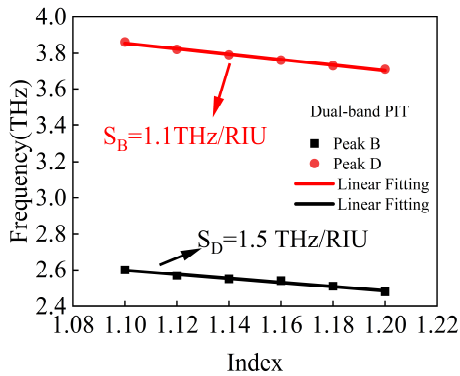

(d)

**Fig. 7 Refractive index sensing performance study. (a) Frequency versus refractive index results for window B (without VO$_2$). (b) Sensitivity of the transparent window at B (without VO$_2$). (c) Frequency versus refractive index results for windows B and D (with VO$_2$). (d) Refractive index sensitivity of transparent windows B and D (with VO$_2$)**

Similarly, the refractive index sensing performance of the dual-band PIT structure is shown in Fig. 7(c) and 8(d). When the refractive index changes from 1.1 to 1.2, the whole transmission spectrum of the dual PIT shifts to lower frequencies as in Fig. 7(c). We define the sensitivity of the transparent window at D as $S_D$. We can see from Fig. 7(d) that the $S_B$ can reach 1.1 THz/RIU which is not much different from the results of single-band PIT. While, the $S_D$ of the transparent window D reaches 1.5 THz/RIU, which indicates that the transparent window at D is more sensitive to changes in refractive index than the transparent window at B under the same conditions. Tab.3 lists the comparison results of the refractive index sensitivity of this design with some of the existing metamaterial structures. The detected results obtained from single window sensing are easily inaccurate due to other disturbances in practical sensing detection. As for the dual-band PIT structure with higher detection accuracy, the two transparent windows have different sensitivities, which benefits us from excluding other interferences against the two resonance peak frequencies in the actual sensing detection. The metamaterial structure presented in this study demonstrates the capability to actively adjust the occurrence of the PIT effect. Additionally, the refractive index sensitivity of both transparent windows exceeds 1 THz/RIU, indicating the potential value of this model in applications related to refractive index sensing.

**Tab.3 Comparison of refractive index sensitivity in different metamaterials**

| Reference | Sensitivity | The number of windows | Year |
| --- | --- | --- | --- |
| [31] | 280GHz/RIU | one | 2021 |
| [32] | 100GHz/RIU | one | 2022 |
| [33] | 875GHz/RIU 775GHz/RIU | two | 2022 |
| [34] | 525GHz/RIU | one | 2023 |
| This work | 1000GHz/RIU 1500GHz/RIU | two | -- |

In summary, we provide a novel metamaterial configuration that exhibits enhanced active tunability by combining Dirac and VO$_2$ materials. In the insulating state, VO$_2$ has a symmetric structure, resulting in a single-band PIT effect. Conversely, in the metallic state, VO$_2$ undergoes an asymmetric structural transformation, leading to a dual-band PIT effect. In addition, the structure achieves multi-band switching modulation through the manipulation of the Fermi energy level of Dirac. This modulation technique allows for a maximum amplitude modulation of up to 98.86%. The resulting structure exhibits a refractive index sensitivity for the clear window that exceeds 1.0 THz/RIU, irrespective of the presence of a VO$_2$ material. Hence, the model exhibits promising possibilities in the realm of optical switch design and applications as a refractive index sensor.

**Ethics declarations**

**Conflicts of interest**

The authors declare no conflict of interest.